\documentstyle[preprint,aps,subeqnarray]{revtex}
\begin{document}
\draft
\preprint{}
\title{Extended technicolor contribution to the $Zbb$ vertex}
\author{Kaoru Hagiwara}
\address{Theory Group, KEK, Tsukuba, Ibaraki 305, Japan}
\author{and}
\author{Noriaki Kitazawa
 \thanks{e-mail: kitazawa@musashi.phys.metro-u.ac.jp}}
\address{Department of Physics, Tokyo Metropolitan University,\\
         Tokyo 192-03, Japan}
\preprint{
\parbox{4cm}{
KEK-TH-433\\
KEK Preprint 95-9\\
TMU-HEL-9501\\
April, 1995\\
\hspace*{1cm}
}}
\maketitle
\begin{abstract}
We show that
 the flavor-diagonal gauge boson of the extended technicolor theory
 contributes with opposite sign to the standard model correction
 for the $Zbb$ vertex.
This mechanism can naturally explain
 the deviation of the LEP result from the standard model prediction
 for the partial width $\Gamma(Z \rightarrow b{\bar b})$.
A smaller value of the QCD coupling, $\alpha_s(m_Z) \simeq 0.115$,
 is then preferred by the $\Gamma(Z \rightarrow \mbox{hadron})$ data,
 which is consistent with both the recent Lattice-QCD estimate
 and the Particle Data Group average.
\end{abstract}
\newpage

The measurement of the $Z$ boson partial width ratio
 $R_b \equiv \Gamma_b / \Gamma_h$ at LEP
 shows a significant deviation
 from the Standard Model (SM) prediction\cite{LEP}.
The measured value $R_b = 0.2202 \pm 0.0020$
 deviates at 2-$\sigma$ level from the SM prediction
 $R_b = 0.2157$ ($m_t = 175$ GeV)\cite{LEP,HHKM}.
The large SM radiative correction proportional to $m_t^2$
 which is specific to the $Zbb$ vertex has not been identified.
Therefore,
 some new contribution to the $Zbb$ vertex
 which can cancel out the SM contribution may be required.

It has been pointed out that
 the ``sideways'' gauge boson of the extended technicolor (ETC) theory
 generates significant correction to the $Zbb$ vertex\cite{chivukula-1}.
The reason is that
 the relatively light (${\cal O}(1)$ TeV) sideways boson
 associated with the top quark mass generation
 should couple with the left-handed bottom quark
 according to the $SU(2)_L$ symmetry.
This contribution is highly model independent.
Flavor-diagonal (``diagonal'') gauge bosons
 appear in the most ETC models,
 and they also contribute to the $Zbb$ vertex\cite{kitazawa}.
The magnitude of the correction
 is comparable with the sideways contribution\cite{kitazawa}
 and the sign is opposite\cite{wu}\footnote{
  In Ref.\cite{kitazawa}
   the sign of the ``diagonal'' ETC boson contribution
   was reported wrongly.
  The error occurred because of the use of an unphysical cut-off procedure.}.
The sideways and the SM contributions make $R_b$ small,
 while the diagonal contribution makes it large.
Therefore,
 if the diagonal contribution is large enough
 to cancel out the other contributions,
 the LEP result can be explained.
In this letter
 we show that this cancellation naturally occurs
 in some models of the ETC theory.
We further note that
 the value of the QCD coupling $\alpha_s(m_Z)$
 as extracted from the $Z$ boson data
 is sensitive to the $Zbb$ correction
 and that the ETC contribution can make its value
 more consistent with both the recent Lattice-QCD evaluation\cite{NRQCD}
 and the global average of the Particle Data Group\cite{PDG}.

Let us consider the one-family-like model
 which was introduced in Ref.\cite{kitazawa}.
The gauge group is
 $SU(N_{TC}+1)_{ETC} \times SU(3)_C \times SU(2)_L \times U(1)_Y$,
 and its fermion contents are
\begin{subeqnarray}
 \left(
  \begin{array}{c}
   \left(
    \begin{array}{cccc}
     U^1 & \cdots & U^{N_{TC}} & t
    \end{array}
   \right)_L
  \\
   \left(
    \begin{array}{cccc}
     D^1 & \cdots & D^{N_{TC}} & b
    \end{array}
   \right)_L
  \end{array}
 \right)
 \quad &\sim& \quad ( N_{TC}+1, \ 3, \ 2, \ 1/6 ),
\\
 \left(
  \begin{array}{cccc}
   U^1 & \cdots & U^{N_{TC}} & t
  \end{array}
 \right)_R
 \quad
 \quad &\sim& \quad ( N_{TC}+1, \ 3, \ 1, \ 2/3 ),
\\
 \left(
  \begin{array}{cccc}
   D^1 & \cdots & D^{N_{TC}} & b
  \end{array}
 \right)_R
 \quad
 \quad &\sim& \quad ( N_{TC}+1, \ 3, \ 1, \ -1/3 ).
\end{subeqnarray}
The lepton sector of the third generation
 and the first and second generations
 are omitted from our discussion for simplicity.
By the breaking of the ETC gauge group $SU(N_{TC}+1)_{ETC}$
 down to the technicolor gauge group $SU(N_{TC})$,
 two kinds of massive gauge bosons are generated:
 massive technicolored sideways gauge boson
 which mediates transition between ordinary quarks and techni-quarks,
 and massive diagonal gauge boson which is flavor-diagonal
 and couples both with ordinary quarks and techni-quarks.

In this naive model
 the masses of the top and bottom quark are degenerate
 for isospin invariant techni-quark condensates,
 $\langle \bar{U} U \rangle = \langle \bar{D} D \rangle$,
 because of the common mass and coupling
 of the sideways boson for each quark.
To be realistic,
 the right-handed top quark and the right-handed bottom quark
 should belong to different representations of the ETC gauge group,
 or a more complicated ETC gauge structure should be introduced.
Instead of considering an explicit ETC model
 that realizes $m_t \gg m_b$,
 we effectively introduce different ETC gauge boson couplings
 for the two right-handed multiplets,
 while keeping the technicolor interaction vector-like.

More explicitly,
 we assign the sideways coupling $g_t \xi_t$ to the left-handed multiplet,
 $g_t / \xi_t$ to the right-handed multiplet with the top quark,
 and $g_t / \xi_b$ to the right-handed multiplet with the bottom quark.
The mass of the top quark is then given by
\begin{equation}
 m_t \simeq {{g_t^2} \over {M_S^2}}
            4 \pi F_\pi^3 \sqrt{{{N_C} \over {N_{TC}}}},
\label{top-mass}
\end{equation}
 where $N_C=3$.
The scale $M_S$ is the mass of the sideways boson and the relation
 $\langle \bar{U} U \rangle \simeq 4 \pi F_\pi^3 \sqrt{N_C / N_{TC}}$
 (from the naive dimensional analysis\cite{manohar-georgi}
  and the leading $1/N$ behavior) is used.
The value of the decay constant $F_\pi$
 in this model with four weak doublets is
 $F_\pi = \sqrt{v_{SM}^2 / 4} \simeq 125$ GeV.
Large top quark mass
 indicates large value of $g_t$ or small value of $M_S$.
The bottom quark mass is given by
\begin{equation}
 m_b \simeq {{g_t^2} \over {M_S^2}} {{\xi_t} \over {\xi_b}}
            4 \pi F_\pi^3 \sqrt{{{N_C} \over {N_{TC}}}}
\end{equation}
 with $\xi_t / \xi_b = m_b / m_t$.
We are assuming that
 the sideways effect can be treated perturbatively,
 and hence we require
\begin{equation}
 {{(g_t \xi_t)^2} \over {4\pi}} < 1
 \quad \mbox{and} \quad
 {{(g_t / \xi_t)^2} \over {4\pi}} < 1.
\label{perturbative-condition}
\end{equation}
The possible range of $\xi_t$ is restricted by this condition.

The couplings of the diagonal ETC boson
 are fixed by the sideways couplings.
For techni-fermions, we obtain the diagonal couplings
 by multiplying the factor
\begin{equation}
 -{1 \over {N_{TC}}} \sqrt{{N_{TC}} \over {N_{TC} + 1}}
\end{equation}
 to their sideways couplings.
For quarks, we obtain them by multiplying the factor
\begin{equation}
 \sqrt{{N_{TC}} \over {N_{TC} + 1}}
\end{equation}
 to their sideways couplings.
These factors
 are determined by the normalization and traceless property
 of the diagonal generator of the ETC gauge group.
The diagonal interaction
 is also chiral in the same way as the sideways interaction.

We now consider the correction to the $Zbb$ vertex.
The sideways boson exchange
 generates the effective four fermion interaction
\begin{eqnarray}
 {\cal L}_{4F}^S &=& - {{(g_t \xi_t)^2} \over {M_S^2}}
                       \left( {\bar Q}_L \gamma^\mu \psi_L \right)
                       \left( {\bar \psi}_L \gamma_\mu Q_L \right)
\nonumber\\
 &=& - {{(g_t \xi_t)^2} \over {M_S^2}}
       \Bigg[
          {2 \over {N_C}}
          \left( {\bar Q}_L {{\tau^a} \over 2} \gamma^\mu Q_L \right)
          \left( {\bar \psi}_L {{\tau^a} \over 2} \gamma_\mu \psi_L \right)
        + {1 \over {2 N_C}}
          \left( {\bar Q}_L \gamma^\mu Q_L \right)
          \left( {\bar \psi}_L \gamma_\mu \psi_L \right)
\nonumber\\
&&
\qquad\qquad\qquad\qquad\qquad
\qquad\qquad\qquad\qquad\quad
        + \left( ({\rm color \ octet})^2 \ {\rm terms} \right)
       \Bigg],
\end{eqnarray}
 where $M_S$ is the mass of the sideways boson,
 $\psi_L \equiv ( t_L \ b_L )^T$ and $Q_L \equiv ( U_L \ D_L )^T$,
 and $\tau^a$ is the Pauli matrix.
Firtz transformation
 for both the Dirac index and the gauge group index
 is performed in the second line.
Below the scale of the technicolor dynamics,
 the techni-fermion currents,
 $J^a_{L\mu} \equiv {\bar Q}_L {{\tau^a} \over 2} \gamma_\mu Q_L$,
 $J_{L\mu} \equiv {\bar Q}_L \gamma_\mu Q_L$, and so on,
 can be replaced by the corresponding currents
 in the low energy effective Lagrangian of the techni-quark sector
\begin{equation}
 {\cal L}_{eff} = {1 \over 4} F_\pi^2 N_C
                  \mbox{tr}
                  \left( (D^\mu \Sigma)^{\dag} (D_\mu \Sigma) \right).
\label{effective-Lagrangian}
\end{equation}
The chiral $SU(2)_L \times SU(2)_R$ symmetry
 for the techni-quark doublet $Q = (U \ D)^T$
 is non-linearly realized in this effective Lagrangian.
The field $\Sigma \equiv \exp (i 2 \Pi / F_\pi)$
 ($\Pi = \Pi^a {{\tau^a} \over 2}$) is transformed as
\begin{equation}
 \Sigma \rightarrow U_L \Sigma U_R^{\dag}
\end{equation}
 corresponding to the chiral transformation
 $Q_L \rightarrow U_L Q_L$ and $Q_R \rightarrow U_R Q_R$,
 where $U_L \in SU(2)_L$ and $U_R \in SU(2)_R$.
The effective Lagrangian is made invariant
 under the local $SU(2)_L \times U(1)_Y$ transformation
 by introducing the covariant derivative
\begin{equation}
 D_\mu \Sigma
  = \partial_\mu \Sigma
  + i g W_\mu \Sigma + i g' B_\mu {{Y_L} \over 2} \Sigma
  - \Sigma i g' B_\mu {{Y_R} \over 2},
\end{equation}
 where $W_\mu = W_\mu^a {{\tau^a}\over 2}$ and $B_\mu$
 are the gauge fields of the $SU(2)_L$ and $U(1)_Y$
 with couplings $g$ and $g'$, respectively.
The fields $\Pi^a$ are the would-be Nambu-Goldstone bosons
 eaten by the gauge fields.
In the unitary gauge, $\Sigma = 1$.

The techni-fermion current $J^a_{L\mu}$ is replaced as
\begin{equation}
 J^a_{L\mu} \longrightarrow
 {1 \over 4} F_\pi^2 N_C
  \mbox{tr}
      \left\{
       - i (D_\mu \Sigma)^{\dag} {{\tau^a} \over 2} \Sigma
       + i \Sigma^{\dag} {{\tau^a} \over 2} (D_\mu \Sigma)
      \right\}.
\end{equation}
In the unitary gauge the third component of the current $J^a_{L\mu}$ is
\begin{equation}
 J^3_{L\mu} \longrightarrow {1 \over 4} F_\pi^2 N_C g_Z Z_\mu,
\end{equation}
 where $g_Z \equiv \sqrt{g^2 + g'^2}$.
Then, we obtain the new $Z{b_L}{b_L}$ coupling:
\begin{equation}
 {\cal L}_{4F}^S \longrightarrow
  - {1 \over 4} {{g_t^2} \over {M_S^2}} \xi_t^2 F_\pi^2 g_Z Z_\mu
    \left( {\bar t}_L \gamma^\mu t_L - {\bar b}_L \gamma^\mu b_L \right)
  + \cdots,
\end{equation}
 and the correction is obtained as
\begin{eqnarray}
 (\delta g_L^b)_{\rm sideways} &=&
 {1 \over 4} {{g_t^2} \over {M_S^2}} \xi_t^2 F_\pi^2 g_Z
\nonumber\\
 &\simeq&
 {1\over 4} \xi_t^2
 {{m_t} \over {4 \pi F_\pi}} \sqrt{{N_{TC}} \over {N_C}} g_Z,
\end{eqnarray}
 where Eq.(\ref{top-mass}) is used in the second line\cite{chivukula-1}.
In the tree level of the SM,
 $g_L^b = g_Z ( -{1 \over 2} + {1 \over 3} s^2 )$
 with $s\equiv\sin\theta_W=g'/g_Z$.

The same technique can be applied
 to obtain the correction due to the diagonal boson\cite{wu}.
The diagonal boson exchange generates the effective four fermion interaction
\begin{equation}
 {\cal L}_{4F}^D = - {1 \over {M_D^2}} J_D^\mu J_{D\mu},
\end{equation}
 where
\begin{eqnarray}
 J_{D\mu} &=& g_t \xi_t \sqrt{{N_{TC}} \over {N_{TC}+1}}
               \left( {\bar \psi}_L \gamma_\mu \psi_L
                    - {1 \over {N_{TC}}} {\bar Q}_L \gamma_\mu Q_L
               \right)
\nonumber\\
          &+& g_t {1 \over {\xi_t}} \sqrt{{N_{TC}} \over {N_{TC}+1}}
               \left( {\bar t}_R \gamma_\mu t_R
                    - {1 \over {N_{TC}}} {\bar U}_R \gamma_\mu U_R
               \right)
\nonumber\\
          &+& g_t {1 \over {\xi_b}} \sqrt{{N_{TC}} \over {N_{TC}+1}}
               \left( {\bar b}_R \gamma_\mu b_R
                    - {1 \over {N_{TC}}} {\bar D}_R \gamma_\mu D_R
               \right),
\end{eqnarray}
 and $M_D$ is the mass of the diagonal boson.
The isosinglet left-handed current ${\bar Q}_L \gamma_\mu Q_L$
 cannot couple to the $Z$ boson,
 but the above effective four fermion interaction
 contains the right-handed current
 $J^3_{R\mu} \equiv {\bar Q}_R {{\tau^3} \over 2} \gamma_\mu Q_R$
 that couples to the $Z$ boson:
\begin{equation}
 {\cal L}_{4F}^D =
  2 {{g_t^2} \over {M_D^2}} {1 \over {N_{TC}+1}}
  \xi_t \left( {1 \over {\xi_t}} - {1 \over {\xi_b}} \right)
  \left( {\bar \psi}_L \gamma^\mu \psi_L \right)
  \left( {\bar Q}_R {{\tau^3} \over 2} \gamma_\mu Q_R \right)
  + \cdots.
\end{equation}
The current is replaced as
\begin{equation}
 J^3_{R\mu} \longrightarrow - {1 \over 4} F_\pi^2 N_C g_Z Z_\mu,
\end{equation}
 and we obtain the correction
\begin{eqnarray}
 (\delta g_L^b)_{\rm diagonal}
 &=& - {1 \over 2} {{g_t^2} \over {M_D^2}} F_\pi^2
                   {{N_C} \over {N_{TC}+1}} g_Z
\nonumber\\
 &\simeq&
     - {1 \over 2} \cdot {{N_C} \over {N_{TC}+1}} \cdot
       {{m_t} \over {4 \pi F_\pi}} \sqrt{{N_{TC}} \over {N_C}} g_Z,
\end{eqnarray}
 where we neglect the small contribution
 which is proportional to $\xi_t/\xi_b$
 and assume $M_D \simeq M_S$.
Therefore,
 the total correction due to the ETC bosons are obtained as
 \footnote{The overall normalization of the correction
           becomes a little smaller,
           if the technicolor dynamics realizes
           large anomalous dimension of the techni-fermion mass operator
           to suppress the flavor-changing neutral current
           \cite{chivukula-2,evans}.}
\begin{equation}
 (\delta g_L^b)_{ETC} =
 \left( \xi_t^2 - {{2N_C} \over {N_{TC}+1}} \right)
 {{m_t} \over {16 \pi F_\pi}} \sqrt{{{N_{TC}} \over {N_C}}} g_Z.
\end{equation}

To analyze the $Zbb$ vertex,
 it is convenient to introduce
 the form factor ${\bar \delta}_b(q^2)$\cite{HHKM}
 in terms of which the $Z{b_L}{b_L}$ vertex function
 is expressed as
\begin{equation}
 \Gamma^{Zbb}_L(q^2) = - {\hat g}_Z
  \left\{ - {1 \over 2} \left[ 1 + {\bar \delta}_b(q^2) \right]
          + {1 \over 3} {\hat s}^2
                        \left[ 1 + \Gamma_1^{b_L}(q^2) \right]
  \right\}.
\end{equation}
The hatted quantities, ${\hat g}_Z$ and ${\hat s}$,
 are the $\overline{MS}$ couplings,
 and the form factor $\Gamma_1^{b_L}(q^2)$ is small in the SM.
The correction due to the ETC bosons is translated as
\begin{eqnarray}
 {\bar \delta}_b(m_Z^2)_{ETC}
  &=& - {2 \over {{\hat g}_Z}} (\delta g_L^b)_{ETC}
\nonumber\\
  &=& \left( {{2N_C} \over {N_{TC}+1}} - \xi_t^2 \right)
      {{m_t} \over {8 \pi F_\pi}} \sqrt{{{N_{TC}} \over {N_C}}}.
\label{correction-ETC}
\end{eqnarray}

The correction within the SM has been estimated.
The one-loop contribution is approximately given by\cite{HHKM}
\begin{equation}
 {\bar \delta}_b^{(0)}(m_Z^2)
  \simeq - 0.00076
         - 0.00217 \left(
                    {{m_t + 36\mbox{GeV}} \over 100\mbox{GeV}}
                   \right)^2.
\end{equation}
The two-loop contribution
 of ${\cal O}(\alpha_s m_t^2)$ is given by\cite{mt-square}
\begin{equation}
 {\bar \delta}_b^{(1)}(m_Z^2)
  = {{\alpha_s} \over \pi} \cdot 2 \left( {{\pi^2} \over 3} - 1 \right)
    {{G_F m_t^2} \over {8 \sqrt{2} \pi^2}}.
\end{equation}
We can neglect the ${\cal O}(m_t^4)$ two-loop contribution
 which is about one order smaller
 than the ${\cal O}(\alpha_s m_t^2)$ contribution.
The total correction within the SM is parameterized as
\begin{equation}
 {\bar \delta}_b(m_Z^2)_{SM}
  = - 0.0099
    - 0.0009 {{m_t-175\mbox{GeV}} \over {10\mbox{GeV}}}
\label{correction-SM}
\end{equation}
 for $\alpha_s = 0.11 \sim 0.12$ and $m_t=(160 \sim 190)$GeV.

{}From the measurement of $R_b$,
 we can obtain the constraint on ${\bar \delta}_b(m_Z^2)$
 without the uncertainty of $\alpha_s$
 and the universal oblique correction\cite{matsumoto}:
\begin{equation}
 {\bar \delta}_b(m_Z^2) = 0.0011 \pm 0.0051,
\label{experiment-Rb}
\end{equation}
 which is about 2-$\sigma$ away
 from the SM prediction (\ref{correction-SM}).
If this deviation is due to new physics,
 the experimental constraint on the new contribution
 to the $Z{b_L}{b_L}$ vertex is
\begin{equation}
 {\bar \delta}_b(m_Z^2)_{new}
  = 0.0110 \pm 0.0051
  + 0.0009 {{m_t-175\mbox{GeV}} \over {10\mbox{GeV}}}.
\label{constraint-on-new}
\end{equation}
There is a 2-$\sigma$ evidence of new physics for $m_t>165$GeV.
If the ETC contribution (\ref{correction-ETC})
 dominates the difference (\ref{constraint-on-new}),
 we find the following constraint
\begin{eqnarray}
 \left( {{2N_C} \over {N_{TC}+1}} - \xi_t^2 \right)
 \sqrt{{{N_{TC}} \over {N_C}}}
 &=& {{8 \pi F_\pi} \over {m_t}}
     \times \left(
             0.0110 \pm 0.0051
             + 0.0009 {{m_t-175\mbox{GeV}} \over {10\mbox{GeV}}}
            \right)
\nonumber\\
 &=& 0.20 \pm 0.09 + 0.005 {{m_t-175\mbox{GeV}} \over {10\mbox{GeV}}}
\label{constraint-on-ETC}
\end{eqnarray}
 where we take $F_\pi=125$ GeV.

The possible value of $N_{TC}$ and the range of $\xi_t^2$ are
 constrained also by the mass formula of the top quark, Eq.(\ref{top-mass}),
 and the perturbative condition, Eq.(\ref{perturbative-condition}).
If we take the ETC scale $M_S \simeq M_D = 1$ TeV,
 the value $N_{TC}=2,3,\cdots,8$ is possible.
The minimal and maximal values of $\xi_t^2$
 allowed by the perturbative condition (\ref{perturbative-condition})
 for $M_S=1$ TeV
 and the experimental constraint from Eq.(\ref{constraint-on-ETC})
 for $m_t=175$ GeV
 are shown in Table \ref{table-4doublet} for several $N_{TC}$ values.
We find that
 the condition (\ref{constraint-on-ETC}) can be naturally satisfied
 in the range $2 \le N_{TC} \le 5$.
It is worth noting here that
 the cancellation between the sideways and the diagonal contributions
 naturally explain the LEP result
 for reasonable range of $N_{TC}$ and $\xi_t^2={\cal O}(1)$.

The one-family model with the small $S$ parameter\cite{Peskin-Takeuchi}
 is proposed by Appelquist and Terning\cite{appelquist-terning}.
In the model
 the techni-lepton condensate largely breaks the weak isospin
 to reduce the $S$ parameter,
 but its scale is small compared with the techni-quark condensate scale
 so that the large weak isospin breaking
 does not affect the weak boson masses, or the $T$ parameter.
To consider the correction to the $Zbb$ vertex in this model,
 we simply change the value of $F_\pi$
 from $F_\pi = \sqrt{v_{SM}^2 / 4} \simeq 125$ GeV
 to $F_\pi = \sqrt{v_{SM}^2 / 3} \simeq 144$ GeV,
 since this model is effectively the three weak doublet model.
If we take the ETC scale $M_S \simeq M_D = 1$ TeV,
 the model with $N_{TC}=2,3,\cdots,20$ is now possible.
The possible range of $N_{TC}$ is extended,
 since the techni-quark condensate is enhanced
 and the ETC gauge coupling becomes small.
The ranges of $\xi_t^2$ for each $N_{TC} \le 8$
 are shown in Table \ref{table-3doublet}.
We find that
 the condition (\ref{constraint-on-ETC}) is satisfied
 in the range of $2 \le N_{TC} \le 7$.

So far we examine constraint in the $Zbb$ vertex
 only from the experiment on the ratio $R_b=\Gamma_b/\Gamma_h$.
In fact the $Zbb$ vertex is constrained
 also by other experiments on the $Z$ pole;
 $\Gamma_Z$, $R_l=\Gamma_h/\Gamma_l$, $R_c=\Gamma_c/\Gamma_h$
 and the peak hadronic cross section $\sigma^0_h$.
There is little sensitivity
 in the forward backward asymmetry $A_{FB}^b$.
It is worth noting that
 except for the ratios $R_b$ and $R_c$,
 all the other observables ($\Gamma_Z$, $R_l$, and $\sigma^0_h$)
 measure just one combination of
 ${\bar \delta}_b(m_Z^2)$ and $\alpha_s(m_Z)$,
 $\alpha'_s = \alpha_s(m_Z) + 1.6 {\bar \delta}_b(m_Z^2)$\cite{HHKM}.
This is because
 the above three accurately measured observables
 depend on $\alpha_s$ and the $Z{b_L}{b_L}$ vertex correction
 only through one quantity, the hadronic width of the $Z$ boson $\Gamma_h$.
As a consequence,
 it has been known\cite{HHKM,Holdom} that
 significant new physics contribution
 to the $Z{b_L}{b_L}$ vertex correction
 affects the $\alpha_s(m_Z)$ value
 extracted from the electroweak $Z$ observables.
Moreover, since the above $Z$ observables
 depend also on the universal oblique correction parameters $S$ and $T$,
 the $\alpha_s(m_Z)$ value extracted from the $Z$ boson data
 should necessarily depend on the three parameters
 $S$, $T$, and ${\bar \delta}_b(m_Z^2)$.
The global fit to extract the value of $\alpha_s(m_Z)$
 has been performed in Ref.\cite{matsumoto}.
In terms of the three charge form factors
 ${\bar g}_Z^2(m_Z^2)$, ${\bar s}^2(m_Z^2)$ and ${\bar \delta}_b(m_Z^2)$
 of Ref.\cite{HHKM}, one finds
\begin{eqnarray}
 \alpha_s(m_Z)
  &=& 0.1150 \pm 0.0044
\label{alphas-fit}
\\
  &-& 0.0032 {{{\bar g}_Z^2(m_Z^2)-0.55550} \over {0.00101}}
   +  0.0015 {{{\bar s}^2(m_Z^2)-0.23068} \over {0.00042}}
   -  0.0042 {{{\bar \delta}_b(m_Z^2)+0.0034} \over {0.0026}}
\nonumber
\end{eqnarray}
 where ${\bar g}_Z^2(m_Z^2)=0.55550 \pm 0.00101$,
 ${\bar s}^2(m_Z^2)=0.23068 \pm 0.00042$, and
\begin{equation}
 {\bar \delta}_b(m_Z^2)=-0.0034 \pm 0.0026
\label{zbb-best-fit}
\end{equation}
 are the best fit values and their 1-$\sigma$ errors.
For a given set of $m_t$ and $m_H$,
 ${\bar g}_Z^2(m_Z^2)$ and ${\bar s}^2(m_Z^2)$ values
 are determined in terms of the $S$ and $T$ values.
The constraint for ${\bar \delta}_b(m_Z^2)$ (\ref{zbb-best-fit})
 has changed from (\ref{experiment-Rb}) by using all the available data.
It should be noted that
 the global constraint (\ref{zbb-best-fit})
 is consistent with the constraint (\ref{experiment-Rb})
 from the $R_b$ data alone,
 while it is still more than 2-$\sigma$ away
 from the SM prediction (\ref{correction-SM}).

The value of $\alpha_s(m_Z)$
 which is obtained from the global fit (\ref{alphas-fit})
\begin{equation}
 \alpha_s(m_Z) = 0.1150 \pm 0.0044
\end{equation}
 is highly consistent with the average value of the results
 given by the Lattice-QCD analyses
 of the bottomonium system\cite{NRQCD}
\begin{equation}
 \alpha_s(m_Z) = 0.115 \pm 0.002,
\end{equation}
 and also with the global average value by Particle Data Group\cite{PDG}
\begin{equation}
 \alpha_s(m_Z^2) = 0.117 \pm 0.005.
\end{equation}

We showed that
 the deviation of the LEP result on $R_b$
 from the SM prediction
 can be naturally explained in the ETC theory.
Since the diagonal and sideways contributions
 to the $Zbb$ vertex are opposite in sign
 and individually larger than the SM contribution,
 the model can naturally explain the 2-$\sigma$ discrepancy
 from the SM prediction
 for reasonable values of $N_{TC}$ and the ETC couplings.
The value of $\alpha_s(m_Z)$ which is extracted from the $Z$ boson data
 becomes small by considering the correction from ETC.
The value is consistent with the recent Lattice-QCD estimate
 and the global average value by the Particle Data Group,
 but is somewhat smaller than that extracted from jet analysis\cite{PDG}.

We are grateful to T.Yanagida and K.Yamawaki for helpful discussions.
We also wish to thank to Y.Sumino
 for the hospitality during our staying at Tohoku university.

\begin{table}[p]
\caption{Possible ranges of $\xi_t^2$ for each $N_{TC}$
         in one-family model, when the ETC boson mass is $1$ TeV.
         The experimental constraint $(\xi_t^2)_{\rm exp}$
         from the $Zbb$ vertex measurement is obtained form
         Eq.(28) for $m_t=175$ GeV.}
\label{table-4doublet}
\begin{tabular}{ccccc}
 $N_{TC}$ & $(\xi_t^2)_{\rm min}$ & $(\xi_t^2)_{\rm max}$ &
 $(\xi_t^2)_{\rm exp}$ & ${{2N_C} \over {N_{TC}+1}}$ \\ \hline
 $2$ & $0.48$ & $2.1$ & $1.8 \pm 0.11$ & $2$   \\
 $3$ & $0.59$ & $1.7$ & $1.3 \pm 0.09$ & $1.5$ \\
 $4$ & $0.68$ & $1.5$ & $1.0 \pm 0.08$ & $1.2$ \\
 $5$ & $0.76$ & $1.3$ & $0.85 \pm 0.07$ & $1$  \\
 $6$ & $0.83$ & $1.2$ & $0.72 \pm 0.06$ & $0.86$ \\
 $7$ & $0.90$ & $1.1$ & $0.62 \pm 0.06$ & $0.75$ \\
 $8$ & $0.96$ & $1.0$ & $0.54 \pm 0.06$ & $0.67$ \\
\end{tabular}
\end{table}

\begin{table}[p]
\caption{Possible ranges of $\xi_t^2$ for each $N_{TC}$
         in the one-family model of Ref.[14]
         with small $S$ parameter, when the ETC boson mass is $1$ TeV.
         The experimental constraint $(\xi_t^2)_{\rm exp}$
         from the $Zbb$ vertex measurement is obtained form
         Eq.(28) for $m_t=175$ GeV.}
\label{table-3doublet}
\begin{tabular}{ccccc}
 $N_{TC}$ & $(\xi_t^2)_{\rm min}$ & $(\xi_t^2)_{\rm max}$ &
 $(\xi_t^2)_{\rm exp}$ & ${{2N_C} \over {N_{TC}+1}}$ \\ \hline
 $2$ & $0.31$ & $3.2$ & $1.7 \pm 0.13$ & $2$   \\
 $3$ & $0.38$ & $2.6$ & $1.3 \pm 0.11$ & $1.5$ \\
 $4$ & $0.44$ & $2.3$ & $1.0 \pm 0.09$ & $1.2$ \\
 $5$ & $0.49$ & $2.0$ & $0.82 \pm 0.08$ & $1$  \\
 $6$ & $0.54$ & $1.9$ & $0.69 \pm 0.07$ & $0.86$ \\
 $7$ & $0.58$ & $1.7$ & $0.60 \pm 0.07$ & $0.75$ \\
 $8$ & $0.62$ & $1.6$ & $0.53 \pm 0.06$ & $0.67$ \\
\end{tabular}
\end{table}

\end{document}